# Automated and Connected Driving: State-of-the-Art and Implications for Future Scenario Analysis

Edgar Jungblut,[a,b,1] Thomas Grube,[a] Jochen Linssen,[a] and Detlef Stolten[a,b]

[a]Forschungszentrum Jülich GmbH, Institute of Energy and Climate Research – Techno-economic Systems Analysis (IEK-3), 52425 Jülich, Germany

[b]RWTH Aachen University, Chair for Fuel Cells, Faculty of Mechanical Engineering, 52062 Aachen, Germany

## Abstract

Automated driving can have a huge impact on the transport system in passenger, as well as freight applications; however, market and technological development are difficult to foresee. Therefore, a systems analysis is called for to answer the question: What is the impact of automated driving on the techno-economic performance of transport systems? It is important to quantify the potential impacts not only on a local scale and for specific use cases but for entire transport systems at large. Here, we provide an overview of the current state of automated driving, including academic research in addition to industrial development. For industrial development, we find that it will take at least until 2030–2040 for automated vehicles to be widely available for passenger transport. For freight transport on the other hand, automated vehicles might already be used within the next years at least on motorways. For academic research, we find that most studies on passenger transport consider shared automated vehicles separated from other transport modes and consider specific regions only. For freight transport we find that operational strategies and usage potentials for level 4 and 5 trucks lack alignment with real-life use cases and driving profiles. Based on this, we develop an analytical framework for future research. This includes a mode choice model for passenger transport demand calculations, a total cost of ownership model for freight trucks, transport statistics for freight flows, a microscopic traffic simulation to assess the impact of automated vehicles on traffic flow, and a road network analysis.

## 1   Introduction

Automated driving has the potential to reduce traffic and make it more efficient on the one hand but also to increase traffic and lead to even more cluttered roads on the other. Furthermore, it might have effects on transport volumes, personal mobility behavior and freight operations, as well as land use and environmental aspects. An overview of some of the affected areas is presented in Figure 1. Understanding the potential impacts of automated driving on the transport system is relevant for a broad range of issues. Answering the question, "What is the impact of automated driving on the techno-economic performance of transport systems?", can also help to answer questions like: How should future cities be designed and what do future mobility concepts look like? What is the potential for road freight operations in comparison to other freight operations? How is automated driving in line with sustainable transportation goals? Many studies on the possible impacts of automated driving in specific contexts have been performed and a lot of work has already been done. The missing step now

---

is to bring all of the results together and develop a system-wide understanding of the potential impacts of automated driving.

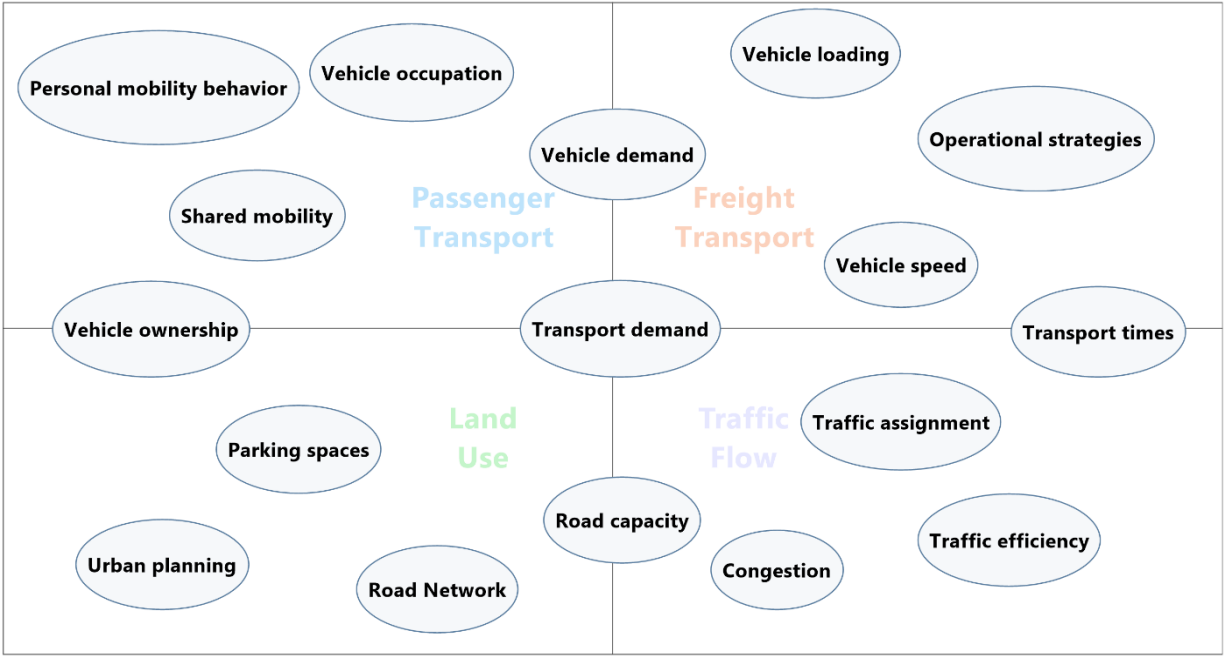

*Figure 1. Selected areas in which automated driving can have an impact on transport (based on Bray and Cebon (2022), Soteropoulos et al. (2019), and Vranken and Schreckenberg (2022)).*

The term "automated driving" is defined by SAE International according to the levels of driving automation schema (SAE, 2021). These levels range from level 0 (No Driving Automation), which means total control by the human driver in any situation, to level 5 (Full Driving Automation), which means that the vehicle can perform all driving tasks on its own, in any situation. In level 1 (Driver Assistance), the driver hands over either the velocity control or steering to the vehicle. One step further, at level 2 (Partial Driving Automation), the driver hands over both the velocity control and steering to the vehicle, but the driver must still supervise the driving tasks throughout. This is relaxed with level 3 (Conditional Driving Automation), in which the driver is allowed to pursue other things in the vehicle besides driving if s/he can take back control over the vehicle within a short period of time (a few seconds) after being notified by it. This will be necessary in cases where the vehicle cannot handle a given situation on its own. At level 4 (High Driving Automation), the driver can focus on entirely different tasks and is not required to take over control within a short period of time. In case the vehicle is not able to handle a particular task, it will transition into a safe position, e.g., driving to the side of the road and stopping there. The levels of automated driving and their requirements are shown in Figure 2.

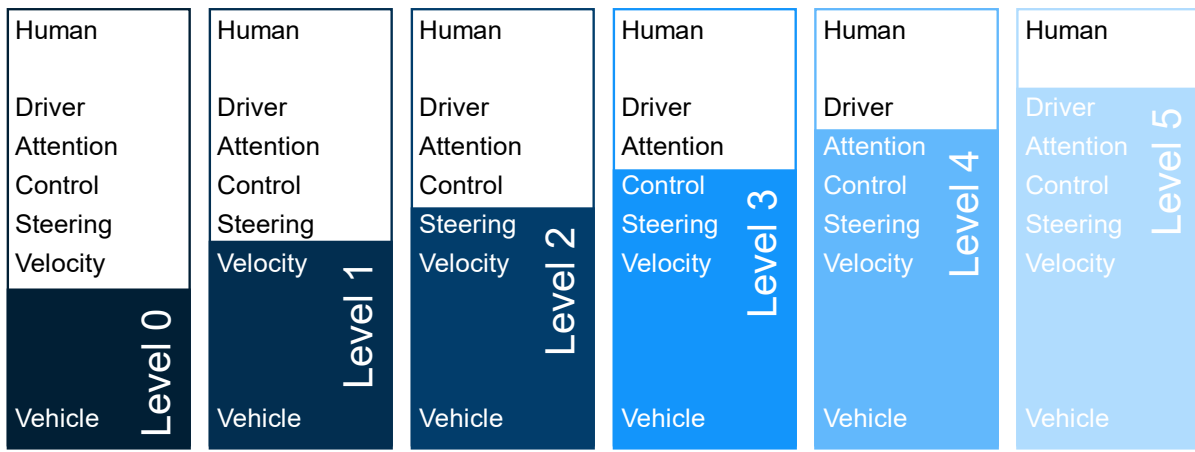

*Figure 2. Levels of driving automation and their requirements as defined by SAE (based on Gaupp and Katzenbach (2019)).*

There have been numerous predictions regarding the advent of true driving automation over the last decade (Hars, 2018). An overview of the prediction and realization dates of different levels of driving automation for passenger vehicles is shown in Figure 3. Some companies, such as Nissan, Audi/Nvidia and Ford have predicted high or full driving automation by the early 2020s (Nissan, 2013; Ross, 2017; Sage and Lienert, 2016). By now, these predictions have been re-adjusted and recent predictions say it will still take at least one or two more decades before fully automated driving will be within reach (Bernhart and Riederle, 2021; Litman, 2021; McKinsey, 2022). Since its inclusion in the DARPA Grand Challenges in 2004 and 2005 and the DARPA Urban Challenge in 2007 (DARPA, 2007), automated driving has come a long way and many engineers, as well as companies, have joined the pursuit of making it viable for day-to-day life (Badue et al., 2021). However, the hurdles still seem to be high, especially for the case of fully-automated driving (Feng et al., 2021).

Nowadays, most car manufacturers have developed advanced driver assistance systems (ADAS) that offer up to automation level 2. Leading the field, Honda and Mercedes have been pushing for the first level 3 vehicles to be produced for the mass market (Edward, 2021; Reuters, 2021). Meanwhile, the autonomous driving technology companies Waymo and Cruise operate fleet cars with level 4 automated driving and have already collected experience kilometers in Phoenix, Arizona, and San Francisco, California (Bellan, 2021a; Hetzner, 2022; Waymo LLC, 2020).

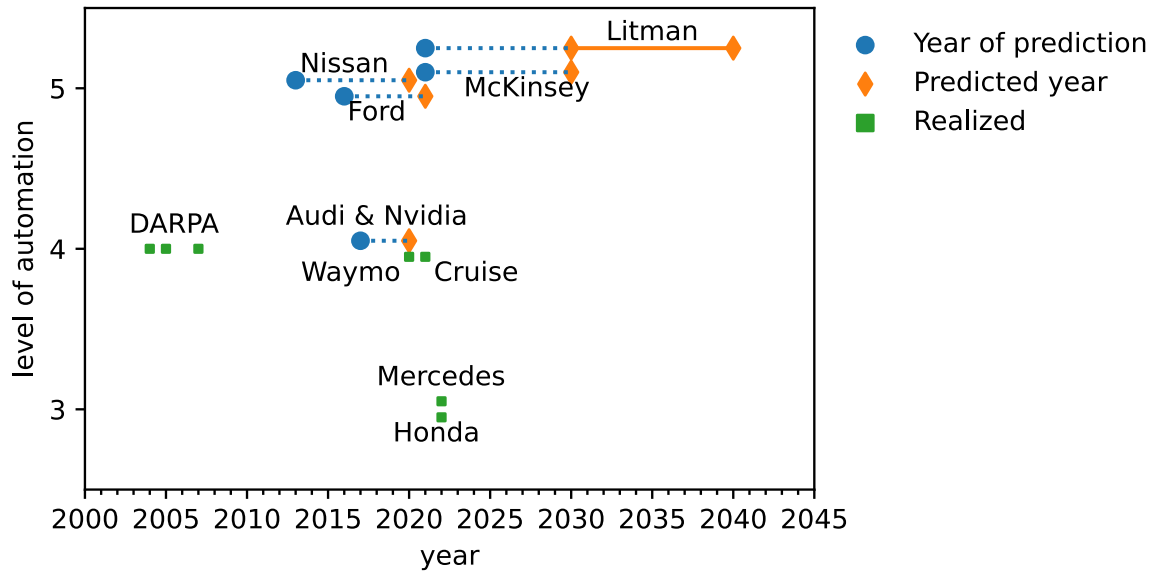

*Figure 3. Timeline of predicted and realized levels of automation for passenger vehicles; see Audi&Nvidia (Ross, 2017), Cruise (Bellan, 2021a; Hetzner, 2022), DARPA (DARPA, 2007), Ford (Sage and Lienert, 2016), Honda (Reuters, 2021), Litman (Litman, 2021), McKinsey (McKinsey, 2022), Mercedes (Edward, 2021), Nissan (Nissan, 2013), Waymo (Waymo LLC, 2020).*

In addition to changing the way that people travel, automated driving also has the potential to have a massive impact on the transportation of goods (PwC, 2018; Roland Berger, 2016). Through automated driving, the costs of transport can be reduced, as there is no longer a need to pay a driver, the door-to-door times of long-distance goods transportation can be reduced when transportation is no longer bound to the permitted driving hours of the human driver, the vehicle can go on driving without having to make any pauses (besides refueling/recharging), and fuel consumption and the emissions of greenhouse gases can potentially be reduced by driving more slowly.

In this study, we provide an overview of the current state-of-the-art in automated driving. Starting with an introduction to the automated driving process, we derive the challenges and potential impacts of automated vehicle deployment. We analyze existing literature on the impacts of automated vehicles, point out gaps therein, and propose guidelines for future analyses as well as a proposal for an appropriate analytical framework.

The remainder of this paper is structured as follows. In chapter 2, we lay out the technical state-of-the-art and current research frontiers in automated driving. We outline the automated driving process in order to understand the challenges and barriers to introduction, as well as estimate future costs. Afterwards, we will give a detailed overview of the literature on the potential impacts of automated driving in chapter 3 to show which aspects have already been investigated. Then, we turn to specific use cases in greater depth. In chapter 4, we will define the requirements and challenges of modeling and analyzing the impacts of automated driving and point out existing research gaps for passenger, as well as freight transport. Drawing on the analyses from the previous chapters, we will then derive guidelines for future research as well as a proposal for an appropriate analytical framework in chapter 5. Finally, in chapter 6 we will summarize our remarks.

## 2 State-of-the-art: Components, costs, and industry activities

In this section, we will introduce the state-of-the-art and current challenges of automated driving to derive information regarding future developments, costs, and deployment timelines. We start with an overview of sensors and algorithms. Then, we look into communication between different vehicles, as well as vehicles and infrastructure. Finally, we introduce the latest industry activities.

## 2.1   Sensors and algorithms

Sensors and algorithms are the key components of automated vehicles. It is necessary to understand their operating principles and limitations in order to assess the possible applications of automated vehicles.

Automated driving can be described as the repeated sequence of the four tasks of sensing, perception, planning, and control (Campbell et al., 2010; Pendleton et al., 2017). This process is visualized in Figure 4. Sensing entails detecting and measuring the environment around the automated vehicle. In the perception step, the sensor data is processed to make sense of the environment. Planning deals with determining the next actions the vehicle should take in order to fulfil its driving task and reach its destination. Finally, control deals with the actual driving task and considers the vehicle's steering and acceleration. We will go through these steps in the following. As the greatest challenges for automated driving are sensing and perception, we will elaborate on these in greater detail.

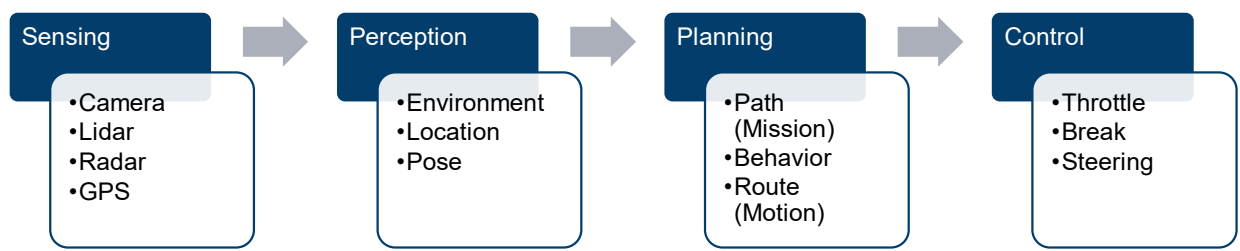

*Figure 4. Automated driving systems architecture (based on Campbell et al. (2010) and Pendleton et al. (2017)).*

### 2.1.1   Sensing and perception

A multitude of sensors form the backbone of each automated vehicle. They are used to perceive the environment around the vehicle and determine its position. This is essential for performing the driving task in a safe and reliable way for the vehicle itself, as well as other traffic participants. The sensors that are most used for automated vehicles are cameras, light detection and ranging (lidar), radio detection and ranging (radar), global navigation satellite systems (GNSS), and inertial measurement units (IMU). Drawing on the works by Liu et al. (2021) and Yeong et al. (2021), we will briefly introduce each of the sensors in the following.

Cameras perceive the environment visually, just as humans do. They are used to identify and classify stationary and moving objects surrounding the vehicle. They have the advantage of being cheap compared to other sensors, costing around 500 USD a piece (Liu et al., 2021). A huge disadvantage of cameras comes, however, from being dependent on good weather and lighting conditions. If the camera vision is somehow obscured, the camera will no longer provide useful information.

Lidar (light detection and ranging) sensors send out laser pulses and measure their reflection. They can detect obstacles and provide information about the distances to these. The lidar sensor data will be represented as a point cloud (coordinates of individual laser pulse reflections). As lidar sensors send out laser pulses on their own, they are not dependent on good lighting conditions. Nevertheless, lidar systems also show decreased performance under bad weather conditions. At the start of automated vehicle development, lidar sensors were fairly expensive, costing up to 75,000 USD a piece (Amadeo, 2017; Domke and Potts, 2020). A lidar sensor costing 90 percent less was then developed by Waymo in 2017 (Amadeo, 2017; Domke and Potts, 2020). By now, lidar sensors are even cheaper, and cost around 1000 USD (Domke and Potts, 2020). In the near future costs might decrease further, to about 700 USD (Watanabe and Ryugen, 2021).

Radar (radio detection and ranging) sensors send out electromagnetic waves and measure their reflection. They can provide information on distances and the relative speeds of obstacles. Radar sensors perform well under any lighting and weather conditions. The main problem of radar sensors is their low spatial resolution compared to cameras and lidar sensors. The costs of radars sensors are about 3000 USD a piece (Liu et al., 2021).

Each sensor has its own functional limitations. In order to ensure proper functionality under many different conditions, the information from all sensors is combined in a process called sensor fusion. Different approaches to this are described by Feng et al. (2021) and Yeong et al. (2021). In essence, the algorithms balance computational complexity, storage requirements, and flexibility in relation to different sensor compositions.

With the setting described until now, the vehicle can perceive its surroundings and therefore maneuver without causing accidents. However, to perform the driving task and plan its route, an automated vehicle needs additional information about its position. The following description of the localization of vehicles is based on the work by Liu et al. (2021).

In order to determine its position in the road network, automated vehicles use street maps and GNSS. Relying on GNSS alone, however, does not permit high enough accuracy, as it usually has an accuracy of a few meters and performs even worse when the signal is obstructed, e.g., by buildings or tunnels (Pendleton et al., 2017). Although this level of accuracy is adequate for a human driver to still perform the driving task properly, for an automated vehicle it is not. Therefore, the information regarding the location of the vehicle must be enhanced. This can be achieved with inertial measurement units (IMU), which measure the vehicle's acceleration and can thereby adjust the positioning as determined by the GNSS system (Liu et al., 2021). Another method makes use of information about the surrounding environment, so-called HD maps, that include not only information about the road network but also visual or point cloud data on the surrounding environment (Liu et al., 2021; Wong et al., 2021). The automated vehicle compares its current sensor data with the sensor data in the map and looks for congruence (Pendleton et al., 2017).

This leads directly to the question of which algorithms are used and what their computational requirements and constraints are.

The research on localization algorithms is focused on enhancing accuracy by using HD maps. The main question here is, which and how much information should be stored in these (HD)maps (Wong et al., 2021). More detailed map data might increase the accuracy, but at the cost of incurring other disadvantages. Firstly, it requires a large amount of memory. Secondly, it is expensive to obtain, as data must be collected for the complete range of sensors. Thirdly, it is even more expensive, as the maps must be constantly kept up-to-date, with Wong et al. (2021) noting, "Prebuilt maps are outdated as soon as they are created." Therefore, a balance between the accuracy and practicability of HD map data must be found.

For the processing of camera, lidar and radar sensor data, a vast number of algorithms have been proposed. Many of these use machine learning or neural networks to detect and classify objects and other traffic participants. The algorithms are trained and tested on large datasets to perform the task properly. It was remarked by Feng et al. (2021) that there are many datasets, but the variety of the traffic situations and weather conditions contained in these datasets may not be very broad. Furthermore, even though the number of datasets available is ever-growing, they will always be fairly limited in comparison to the vast amount of possible traffic and weather conditions automated vehicles may face in the real world (Uricár et al., 2019). It must be ensured that the automated vehicle can handle every situation it may face. It is questionable whether this is possible with ever more data or if another form of validation is needed (Campbell et al., 2010).

To perform the driving task safely and prevent accidents and potential fatalities, the automated vehicle needs to be able to respond to changes in outer circumstances within a short period of time (Feng et al., 2021; Lin et al., 2018; Tabani et al., 2021). This creates the need for fast computational systems. Different algorithms and architectures that offer solutions to this constraint are analyzed by Lin et al. (2018). They compare the computational time of the algorithms to a threshold of 100 ms to ensure the system reacts at least as fast as a human would. The authors find that object detection, tracking, and localization take up about 94% of the computational time.

The most critical aspect of the algorithms, besides producing results in a timely manner, is that the produced results must be reliable (Lin et al., 2018; Tabani et al., 2020). Furthermore, the processing of sensor data must not consume too much power (Lin et al., 2018). This constraint is especially relevant because automated vehicles are probably also going to be electric. For these, a computational system with a 1.8 kW power demand may decrease the overall driving range by up to 12% (Lin et al., 2018).

### 2.1.2 Planning and control

We do not go into much detail regarding the planning and control processes, as these do not constitute the main challenges of automated vehicle development. This short explanation is intended to complete the understanding of the automated driving process. For more detailed information about the processes and algorithms used, see Pendleton et al. (2017) and the references cited therein; the following summarizes remarks from that study.

Planning can be separated into three steps, namely: mission, behavioral, and motion, representing long-, medium-, and short-term planning. Mission planning is concerned with designing a path from the current vehicle position to its desired destination. To perform this task, the vehicle uses data from its GNSS system and available maps of the road network. At the next level, behavioral planning deals with the present driving situation and concerns abiding by traffic rules and interactions with other traffic participants. At the last level, motion planning entails performing the actual driving task moment to moment, i.e., preventing collisions and accidents.

Based on the information given by the planning algorithms, the vehicle accelerates and steers. These processes are regulated by feedback and feedforward control systems. The actual velocity and steering angle are compared to the desired ones and constantly adapted.

## 2.2 Communication

After considering the operating principles of individual automated vehicles, we will now turn to the implications of sharing information with other vehicles and the infrastructure. Even though automated vehicles are expected to function on their own, communication between them and with the infrastructure may offer benefits to the individual vehicles, as well as the transport system. These benefits, however, come with additional costs for equipping the vehicles and infrastructure for communication and must be weighed against these. In the following, we will present the three most significant use cases of communication.

First, shared sensor and environment data would provide vehicles with a better understanding of the surrounding situation (Dong et al., 2020; Pendleton et al., 2017). This might improve safety especially in city driving in which many obstacles make it nearly impossible for a single vehicle to take in the entire traffic situation.

Second, vehicle-specific information can be shared between vehicles. This could be information about velocities and accelerations, as well as planned routes and near-term behavior that would enhance the driving behavior of automated vehicles (Fagnant and Kockelman, 2015; Pendleton et al., 2017). A prominent example for connected driving is platooning, whereby vehicles can follow each other with short safety distances because they know about the other vehicles' (planned) driving behavior, including accelerating and braking (Fagnant and Kockelman, 2015). Platooning has the potential to reduce fuel consumption through more efficient driving and lower drag, especially in the context of (long distance) freight transport (Fagnant and Kockelman, 2015). We will look at this in more detail in chapter 3.2.

Third, infrastructure can provide data about the traffic situation in certain areas. This would enable vehicles to adjust their routes to traffic volumes and travel times to make smarter routing decisions (Liu et al., 2021).

As mentioned before, equipping the vehicles and infrastructure for communication comes with additional costs. Multiple studies therefore deal with the optimal placement of roadside units (RSUs) for vehicle-to-everything (V2X) communication. The proposed algorithms perform a cost-optimization under constraints for latencies and coverage. Moubayed et al. (2020) assumed monthly operational costs of 7,500 USD for RSUs with cloud-based services, and 15,000 USD for those in which the services take place themselves. Furthermore, safety-critical communication will always be performed with the latter because of strict latency constraints (Moubayed et al., 2020). For the correct notification of emergency services in cases of accidents, Barrachina et al. (2013) proposed that at least four RSUs are needed per square kilometer, with a uniform mesh distribution for low vehicular density areas ($< 70$ veh./km$^2$) and a lower number of RSUs for high vehicular density areas ($> 70$ veh./km$^2$), as a larger number of vehicles are available for vehicle-to-vehicle (V2V) communication and fewer access points to vehicle-to-infrastructure (V2I) communication are needed. For Germany with an area of 357,592 km$^2$ (Destasis, 2023) this would amount to total costs of 10.7–21.5 billion USD.

## 2.3  Industry activities

We will now look at what is happening outside of academia in the realm of the research and development of car manufacturers and autonomous vehicle development companies. This will provide an understanding of the current state and expected timelines for the deployment of automated vehicles in the mass market.

It is noteworthy that in the early stages of automated driving research, efforts were not driven by traditional car manufacturers but rather technology companies and start-ups. The most prominent examples include Google's self-driving car project (which was spun off and became an independent firm called Waymo), the mobility-as-a-service provider Uber (which has now given up on developing autonomous vehicles and sold its division to Aurora (Metz and Conger, 2020)) and the self-driving car company, Cruise. The traditional car manufacturers are starting to join the race for automated vehicles by cooperating and acquiring start-ups that develop automated driving technology (Badue et al., 2021). This has been fostered by the enormous costs and safety requirements the development of a vehicle requires (Metz, 2021). The first traditional car manufacturers to announce commercially-available automated vehicles up to level 3 in the next years were Honda and Mercedes (Edward, 2021; Reuters, 2021). In contrast, the automated vehicle development companies Waymo and Cruise are already operating level 4 automated vehicles (Hetzner, 2022; Waymo LLC, 2020). As already discussed earlier, an overview of the prediction and realization dates of different levels of driving automation for passenger vehicles is shown in Figure 3.

Up until now, we have only considered passenger transport, but there is a strong case to be made for automation in freight transport. We will introduce it here and discuss it further in chapter 3.2.

An overview of the funding raised by companies developing automated trucks is presented in Figure 5. By far the two largest of these are Waymo and Aurora, followed by Inceptio Technology, TuSimple, Plus, Embark Trucks, Kodiak Robotics, and Gatik AI (Crunchbase, 2022a-h). All of the leading companies employ combinations of cameras, lidar, and radar (Aurora Innovation Inc., 2022a; PlusAI Inc., 2022a; TuSimple Holdings Inc, 2022a; Waymo LLC, 2022). The deployment scenarios for level 4 automated trucking given by the companies fall within the next years. Plus has already started to deliver its first commercial product, still relying on a driver at first but with the intention of soon moving to level 4, without a driver (PlusAI Inc., 2022b, 2021). TuSimple plans to deliver commercially-available level 4 automated trucking by 2024, after their first test run without a driver in December 2021 (Bellan, 2021b). Aurora intends to deploy the first commercial level 4 automated trucks by 2023 in selected states of the US and expand to all states over the course of eight more years (Aurora Innovation Inc., 2022b, 2021).

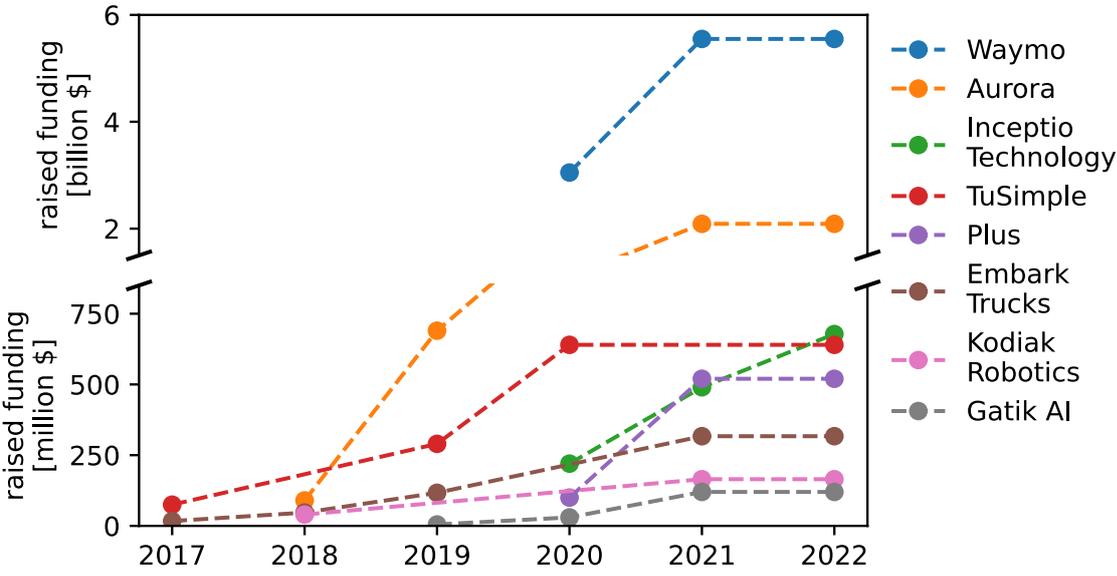

*Figure 5. Total funding raised by autonomous truck companies based on (Crunchbase, 2022a-h).*

To conclude the remarks of this chapter: Automated vehicles rely on different sensors that allow them to perceive their environments to make driving decisions. These sensors will add considerable costs to them, but these additional costs will likely decrease in the future due to maturing technology and mass production. Research on automated vehicles is primarily concerned with improving environmental perception. In particular, the high variety of traffic situations and environmental conditions an automated vehicle may encounter poses a major challenge. To ensure safety, perception must be accurate while the computational times should be low. Furthermore, the algorithms must not consume too much power, as this would limit the operational range of the vehicles. Automated vehicles are intended to function entirely on their own. Nevertheless, communication with other vehicles and infrastructure might improve their environmental perception and may also foster better traffic management. Currently, the automated vehicle development companies Waymo and Cruise lead the field with level 4 automated vehicles operating in fleet services in selected US cities, whereas traditional car manufacturers are a little behind in development but have started to enter the race by acquiring start-ups in the field. It can be expected that automated vehicles for passenger transport will not be available to the general public in most areas before 2030–2040. In the area of freight transport, on the other hand, level 4 automation for highway operations is expected to be commercially-available within the next years.

# 3 Literature overview: Deployment scenarios for automated vehicles

We will now look at research on the impact of automated vehicles that has been conducted until now. Since the use of automated road vehicles in the real world has so far been limited to small areas of application and tests, current research mainly consists of modeling-based studies. Here, we focus on the considered regions and levels of automation as well as the assumptions and methods used. We outline the state-of-the-art of research in this chapter and use that information in chapter 4 to identify where gaps might exist. We will first deal with automation in passenger transport and then consider automation in freight transport.

## 3.1 Passenger transport

Automated vehicles may bring about changes in passenger transport in two respects. First, higher costs of vehicles may lead to a decrease in privately-owned vehicles and promote new mobility concepts like car- (where people share a vehicle and use it for trips in succession) or ride-sharing (where people share a vehicle and use it for trips together if trips have a large part in common). The market introduction of highly automated vehicles is therefore expected to be undertaken via shared vehicle and fleet services (Narayanan et al., 2020). Second, automated vehicles enable the driver to use the time in the vehicle for other tasks, which would make this more time-efficient. In transport modeling, travel cost and time are considered when deciding on trips and transport modes. To this end, travel time is converted into a cost equivalent via the value of travel time, an amount of money to compensate for the time lost traveling. This value is highly individual for people and trip purposes and might fall substantially for automated vehicles, which would lead to the promotion of longer trips. Changes in passenger transport are assessed by changes in overall vehicle kilometers travelled (VKT), the number of vehicles needed, and related effects.

A review of available modeling studies on automated driving impacts between 2013 and 2018 was performed by Soteropoulos et al. (2019). The authors found that most of the available literature on the impact of automated driving focuses on high levels of automation, i.e., vehicles that can handle even complex scenarios like driving in urban environments. This might be because high levels of automation may substantially change passenger transport with new mobility concepts emerging and new user groups gaining access to them. Furthermore, the focus is often on shared vehicles rather than privately-owned ones.

Studies of privately-owned automated vehicles show a strong increase in overall vehicle kilometers travelled (VKT). In a study for the city of Delft in the Netherlands using a mode choice model, it was found that when replacing privately-owned vehicles with automated ones, the share of car trips rose from 43.6% to 47%, and VKT increased by 17% (Correia and van Arem, 2016). When further assuming that the value of travel time in automated vehicles is only half as high as for conventional cars, the car share rose to 53.4% and VKT by 49% (Correia and van Arem, 2016). If, furthermore, automated vehicles only park in outer city regions, VKT drastically increased, by a factor of 2.9, with 62% of these being empty kilometers traveled by the vehicles (Correia and van Arem, 2016). In another modeling study that analyzed the influence of parking price schemes for work trips, it was found that replacing conventional cars with automated ones that can drive long distances to find a parking space either in rural areas or at home would increase vehicle travel times by about 50% and therefore may cause congestion (Kang et al., 2022). To counter the effects of increased VKT due to empty rides in automated vehicles, the introduction of a $5/h fee for empty vehicle driving was found to reduce VKT in car routing and parking choice simulations for the city of Toronto by 3.5% (Bahrami and Roorda, 2022). In a study of the Chicago metropolitan area using the POLARIS modeling framework, an increase in VKT of up to 43% for the use of level 4 privately-owned automated vehicles was found when all conventional cars were replaced by automated vehicles and the value of travel time was assumed to be half as high as

for conventional cars (Auld et al., 2018). In a further study of the same region using the POLARIS framework, a 3% VKT increase for an automated vehicle penetration rate of 20% and a value of travel time reduction of 25% for these was found (Auld et al., 2017). Furthermore, a 79% VKT increase for the complete penetration of automated vehicles, a value of travel time reduction of 75%, and a 77% road capacity increase was found (Auld et al., 2017).

In contrast to privately owned automated vehicles, a multitude of the modeling studies assume that highly automated vehicles (levels 4 and 5) are not owned by individual people anymore, and that instead vehicles are shared between people or operated within fleets (Soteropoulos et al., 2019). There are two concepts for sharing. First, people can share vehicles, making trips in succession of one another. Second, they can share rides, making their trips together when they have a similar route. These two concepts utilize vehicles much more than current privately-owned ones, which are not used for more than 20 hours each day (Hancock et al., 2019; Martínez-Díaz and Soriguera, 2018).

For a fleet of shared vehicles in a hypothetical American urban area of 10x10 miles, it has been shown in modeling studies that one shared automated vehicle (without ride sharing) may replace up to eleven conventional, privately-owned vehicles when considering the demand for short trips of less than 15 minutes (Fagnant and Kockelman, 2014). In a study of the area of Zurich, similar results were found, with a single shared automated vehicle being able to replace up to ten conventional privately-owned ones if 10% of conventional car trips were assumed to be served by automated vehicles and waiting times of up to ten minutes were accepted (Boesch et al., 2016). Again, a similar picture emerges from studies of the city region of Berlin, in which car trips within the city were replaced by automated vehicle trips and the fleet size of the required automated vehicles to serve the demand was 10% of the conventional car fleet size, with capacity shortages only arising during peak hours in the afternoon (Bischoff and Maciejewski, 2016). The authors further argue that the fleet size could even be smaller if it did not have to account for the peak hours, as the automated vehicles of such a fleet are busy for only 7.5 hours per day and idle for the remainder of it (Bischoff and Maciejewski, 2016).

The reduction of cars needed might seem like a good thing in view of making transport more sustainable. However, there are other factors to consider that show that the effects of shared automated vehicles might still warrant further research.

First, the need for a smaller vehicle fleet does not mean that fewer vehicles are going to be produced, as vehicle lifetimes will be reduced due to the stronger utilization of them, and so the number of vehicles produced may remain the same overall (Fagnant and Kockelman, 2014). To assess this claim, we will perform a brief analysis of vehicle production rate dependencies in Apendix A.

Second, the important measure to consider when assessing sustainability from a GHG perspective is the overall vehicle kilometers travelled (VKT). For the above-mentioned study on the hypothetical American urban area, an overall VKT increase of 11% for shared automated vehicles was found (Fagnant and Kockelman, 2014). In a modeling study based on real-world data from the urban area of Langfang, China, a VKT increase of 12% was found when conventional car trips were replaced by shared automated vehicles without ride-sharing (Liu et al., 2022). However, it is further shown that the overall VKT can decrease when the option for pairwise ride sharing (sharing rides in pairs of two) is introduced. If 60% of passengers participate in pairwise ride-sharing, the VKT is 2% lower than for the base case with conventional cars and if 100% of passengers participate in pairwise ride-sharing, a decrease of 14% in VKT is found (Liu et al., 2022). In a study of the Tel Aviv metropolitan area, where all vehicular travel inside this area was replaced by shared automated vehicles, VKT could be reduced by 20% through ride-sharing with the average vehicle occupancy being 2.1, and therefore almost doubled compared to normal standards of 1.1 for conventional cars today in Tel Aviv (Ben-Dor et al., 2019).

The usage of automated vehicles may also have impacts on the design of future cities. First, there may not need to be as many parking spaces in city centers. In an analysis of parking spaces for a hypothetical US city where conventional vehicles were replaced by shared automated vehicles (with and without ride-sharing), it was found that up to 90% of the parking spaces in urban areas might be obsolete (Zhang et al., 2015). Further implications of automated vehicles may be that people will think differently about their places of residence. If commuting with automated vehicles is more pleasurable, as suggested by Kolarova et al. (2019), longer commuting distances might be accepted by users, and therefore they will not need to live as close to their workplace and might consider moving to suburbs and rural areas (Haboucha et al., 2017). In a study using a multinomial logit model for housing choice, a slight move away from the workplace was found that would lead to an increase in overall VKT, at least for commuting (Zhang and Guhathakurta, 2021).

A further benefit of (shared) automated vehicles lies in an increased accessibility of regions that are not as well-connected to public transport. In a study of municipalities in Switzerland using the Swiss national transport model, it was found that the car accessibility of extra-urban areas measured by travel time may increase in the face of (shared) automated vehicles entering the market, whereas the impact on cities will be either weak or even negative due to the increased traffic and congestion; lastly, remote alpine regions were found to be unaffected (Meyer et al., 2017).

A more far-reaching review of the potential impacts of shared automated vehicles was performed by Narayanan et al. (2020). In this study, the impacts were delineated into the seven categories of traffic and safety, travel behavior, economy, transport supply, land use, environment, and governance.

All studies presented rely on people's willingness to use automated vehicles and take part in mobility concepts like car- and ride-sharing. For the latter, assumptions can be based on the popularity of car- and ride-sharing service operating right now (realized preferences), but the current form of car- and ride-sharing might be quite different from the future one, consisting of shared automated vehicles. Thus, people's acceptance and use of such modes might be different in the end.

In a mode choice study for a synthetic city resembling Cologne (Germany), calibrated to German mobility data from 2017, shared automated vehicles were not highly utilized (0.4% over all scenarios) and even after banning conventional private vehicles with a modal share of 31.9%, the modal share of shared automated vehicles only rose to 4.8%, with most passenger kilometers now taken over by public transport (Reul et al., 2021).

To make up for the differences from current vehicles, many studies have been conducted on the attitudes of people towards the use and perceived benefits of automated vehicles, as well as car- and ride-sharing concepts for these. However, the studies are all based on the hypothetical choices of people (stated preferences), as automated vehicles are not yet available. This has the drawback that the studies may not reflect the actual behavior of the people when the technology is in fact available. Nevertheless, we want to share some of the findings here.

First, it should be noted that there is no strong basis for knowledge regarding the potential user groups of shared automated vehicles and it is unclear if and how elderly or younger people would use them, as these user groups are highly heterogenous (Krueger et al., 2016). In a 2017 study, survey respondents were found to be skeptical towards automated vehicles, with 25% of them not wanting to use shared automated vehicles, even when they were completely free of charge (Haboucha et al., 2017). Furthermore, in a 2016 study, it was found that people do not perceive the advantages of performing non-driving activities while spending time in automated vehicles like working or using their phones (Yap et al., 2016). In a 2020 study, it was found that about 30% of people are not willing to pay anything extra for automated vehicles of level 3 or higher compared to conventional vehicles, with only 20% being willing to pay 10,000 USD or more extra (Elvik, 2020). The authors of the study conclude that this is an indicator that automated vehicles will be too expensive for private ownership for most people at first, as they estimate automated vehicles to cost 10,000–40,000 USD more than conventional vehicles (Elvik, 2020). This would favor an introduction of automated vehicles through shared vehicles and fleets. People who have experience with car-sharing and already use it today were found to be more willing to use shared automated vehicle services (Krueger et al., 2016; Nazari et al., 2018).

Most studies assume different states and market penetration levels of automated driving for their scenarios without any time horizon connected to them. The recent notion is that fully automated driving, and so the modes described in the presented studies, is still far away. It will take at least 10–20 more years until automated driving technology is ready for the mass market (Litman, 2021). It should be noted that it will take even more time for automated vehicles to make up a substantial share of all vehicles on the road, as vehicle lifetimes are long and new technologies therefore take time until they are widely applied in all vehicles (Litman, 2021). Assuming a 100% technology adaptation for new vehicles in the US market with an average vehicle lifetime of 16.6 years, it would take 19.6 years until the new technology would be present in 90% of the vehicle fleet (Keith et al., 2019).

In this section, we outlined the state-of-the-art of research in passenger transport. To conclude our remarks, Figure 6 summarizes the research scope of the presented modeling studies regarding considered modes and regions. The studies often consider automated vehicles separated from other transport modes, especially when analyzing new mobility concepts such as shared automated vehicles. Furthermore, the studies are often restricted to narrow regional areas of application, such as urban or metropolitan settings.

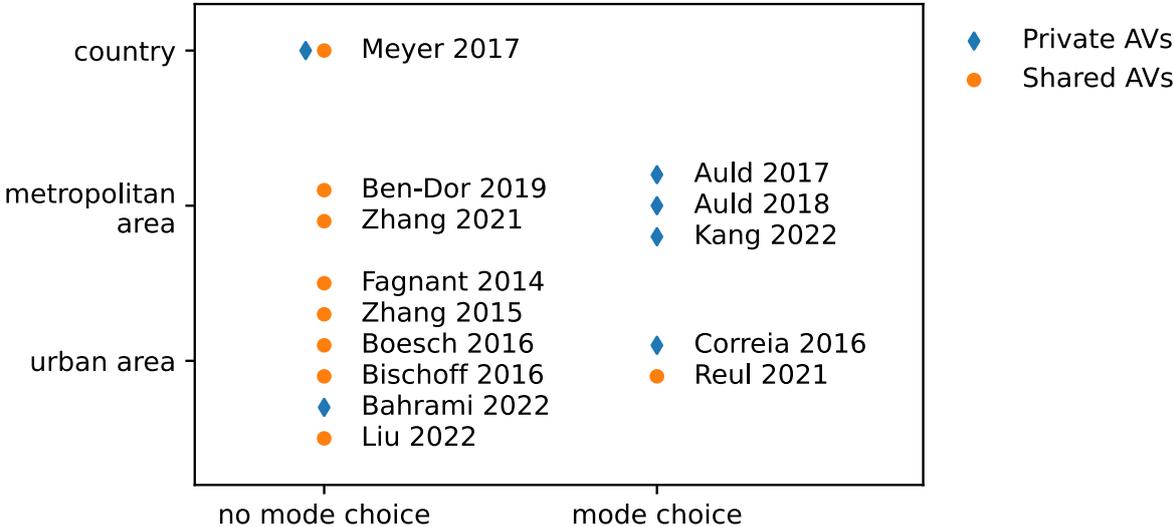

*Figure 6. Research scope of the cited papers analyzing the influence of automated vehicles on mode choice and VKT regarding considered modes and regions; see: Auld 2017 (Auld et al., 2017); Auld 2018 (Auld et al., 2018); Bahrami 2022 (Bahrami and Roorda, 2022); Ben-Dor 2019 (Ben-Dor et al., 2019); Bischoff 2016 (Bischoff and Maciejewski, 2016); Boesch 2016 (Boesch et al., 2016); Correia 2016 (Correia and van Arem, 2016); Fagnant 2014 (Fagnant and Kockelman, 2014); Kang 2022 (Kang et al., 2022); Liu 2022 (Liu et al., 2022); Meyer 2017 (Meyer et al., 2017); Reul 2021 (Reul et al., 2021); Zhang 2015 (Zhang et al., 2015); and Zhang 2021 (Zhang and Guhathakurta, 2021).*

Before pointing out potential research gaps for passenger transport in chapter 4, we will cover the state-of-the-art of research in freight transport in the next section.

## 3.2   Freight transport

In comparison to passenger transport, automation in freight transport is much less addressed in academic research. This may be due to the fact that there do not seem to be as many usage scenarios for and factors influencing the adoption of automated driving in freight transport. Automated driving in freight transport is not a question of personal preferences, but an economic one. If large benefits can be achieved by automated driving, it will be adopted quickly (Engström et al., 2019); therefore, freight transport might be the first area in which automated driving will be established.

The technological requirements are similar to those in passenger transport, but the automation of freight transport is supposed to be simpler than that of passenger transport, because (long-range) trucks travel on highways most of the time, which is a much simpler traffic environment (Engström et al., 2019; TuSimple Holdings Inc, 2022b). The interactions with other participants are limited in this environment, as there are no cross sections, pedestrians or cyclists, and all vehicles are traveling in the same direction. Nevertheless, freight transport operations are not restricted to highways. Driving on non-highway roads will most likely not be automated at first and therefore a driver will still be needed for these settings. In the following, we will discuss the different cost-saving potentials of automated freight transport analyzed so far.

The area that has gained the most attention in freight transport research (even before automation) is the effect of platooning on fuel savings. Platooning and fuel savings can already be achieved by low levels of automated driving, i.e., automated distance between vehicles (Roland Berger, 2016). In different studies, it was found that fuel savings achieved through the platooning of two trucks were around 4% to 15% dependent on the distance between the vehicles and their loading (Slowik and Sharpe, 2018; Tsugawa et al., 2016). However, platooning is only possible in a restricted part of the trip. In a study of trucking miles in the United States, it was found that about 66% of them are suitable for platooning when assuming that it is possible if the vehicle velocity is above 50 mph (80 km/h) for at least 15 minutes (Muratori et al., 2017). For long-distance driving, the share of platoonable miles even increases to 77% (Muratori et al., 2017). In combination with fuel savings through platooning of 6.4% (Lammert et al., 2014), the overall fuel savings were estimated to be 4% (Muratori et al., 2017). As fuel amounts to 25% to 40% of the total cost for long-distance trucking (Sharpe, 2017), the overall cost may decrease by 1% to 2% through platooning. Beyond that, there are operational challenges to it, e.g., freight trips must be coordinated to make the best use of platooning.

As a further option, reductions in driving speeds have been considered lately. Through extended driving hours, automated trucks can travel long distances in shorter times than human-driven ones. Portions of the time gained could be spent on reducing driving speeds and therefore reducing fuel consumption. A target speed reduction from 90 km/h to 70 km/h for a tractor-trailer of 29.5 t was found to decrease fuel consumption by 26% in a drive cycle simulation (Bray and Cebon, 2022). However, this was found to lead to a cost reduction only when driver costs were eliminated, since the additional driver wages through longer driving hours outweighed the cost savings for fuel (Bray and Cebon, 2022).

A much larger cost benefit for freight operations may come from eliminating the need for a driver in levels 4 and 5 automated trucks (Roland Berger, 2018). Such trucks do not need to make any stops for driver rest times and their utilization rate can therefore be increased from 29% to 78% (PwC, 2018), which could reduce delivery times through longer possible driving hours. The largest potential for cost reductions, however, comes from the reduction of labor costs for drivers, as these may go down by up to 90% (Roland Berger, 2016) and make up a share of about 80% of total savings (PwC, 2018).

Furthermore, the current shortage of truck drivers (BAS, 2021; ONS, 2021) may be relieved, even though this is not a short- or medium- but rather a long-term solution (BMVI, 2020).

Automated truck operations are expected to be limited to highway operations at first, as these environments are simpler than off-highway ones (Engström et al., 2019; TuSimple Holdings Inc, 2022b). A business model would therefore make use automated trucks on highways and conventional trucks for off-highway operations (Engström et al., 2019; Roland Berger, 2018). The overall operating cost reductions for this case are highly dependent on the trip length and share of automated driving for the trip, e.g., for a trip of 800 km with 480 km being operated by an automated truck, overall savings of 22% might be achieved (Roland Berger, 2018). Estimates for overall operational cost reductions for level 5 automation (100% of the trip being automated) are in the range of 30% (ITF, 2017) to 47% (PwC, 2018). Such cost reductions might as well influence the demand for road freight transport. In a study for the United States using a random utility-based multiregional input-output model, a per ton-kilometer cost reduction for trucks of 25% was found to lead to an 11% increase of ton-kilometers for trucks, whereas rail ton-kilometers were found to decrease by 4.8% (Huang and Kockelman, 2020).

Automated trucks have large investment costs. One prognosis estimated the overall additional cost of full automation to be 23,400 USD (Roland Berger, 2016). A later prognosis estimated the additional cost of level 4 trucks at about 30,000 USD initially in 2025, and 15,000 USD by 2030 due to scaling effects (Roland Berger, 2018). Other studies have estimated the costs to be nearly double the initial 23,400 USD (Slowik and Sharpe, 2018). In a 2018 study, industry expectations for the deployment of level 4 automation were found to be 4–10 years and 7–20 years, respectively, for level 5 automation (Slowik and Sharpe, 2018).

In this section, we outlined the state-of-the-art of research in freight transport. In conclusion, research on automated freight transport until now has considered not only full automation but also cases in which automated trucks can only operate in restricted areas, as this will most likely be their state during the early years. Table 1 summarizes the research scope of the presented studies. It shows that cost studies for automated freight transport have been performed by a wide range of organizations. The key factors for cost savings are seen in fuel savings (via platooning or lower driving speeds), the omission of driver costs, and the associated gains in operational times. Concrete business models and operational strategies are proposed mostly by consultancies.

*Table 1. Research scope of the cited papers analyzing automated freight vehicles; see: Bray 2022 (Bray and Cebon, 2022); Engström 2019 (Engström et al., 2019); Huang 2020 (Huang and Kockelman, 2020) ITF 2017 (ITF, 2017); Lammert 2014 (Lammert et al., 2014); Muratori 2017 (Muratori et al., 2017); PwC 2018 (PwC, 2018) Roland Berger 2016 (Roland Berger, 2016); Roland Berger 2018 (Roland Berger, 2018); Sharpe 2017 (Sharpe, 2017) Slowik 2018 (Slowik and Sharpe, 2018); Tsugawa 2016 (Tsugawa et al., 2016)*

| Reference | Type of study | Research Scope: Operational Strategy, Cost, Fuel Savings |
|---|---|---|
| Bray 2022 | Driving cycle energy demand simulation and operational cost assessment | Costs: Vehicle, Maintenance, Fuel, Driver, Freight value of time<br>Fuel Savings: Driving cycle speed reduction |
| Engström 2019 | Conference breakout session | Operational Strategy: Identified need for use case specific operational strategies |
| Huang 2020 | Multiregional input-output model including freight mode choice | Costs: Mode shift to road freight transport caused by reduced per ton-kilometer costs for automated trucks |
| ITF 2017 | Literature review on impact of automated driving on labor market | Costs: Operational cost including driver and fuel |
| Lammert 2014 | Truck platooning experiment | Fuel Savings: Two trucks with different platooning settings (weight, distance, speed) |
| Muratori 2017 | Statistical analysis of truck trip data | Fuel Savings: Estimation of platoonable miles |
| PwC 2018 | Operational cost assessment | Operational Strategy: Level 5 automated supply chain<br>Costs: Operational cost including driver, cabin, and sensors |
| Roland Berger 2018 | Operational cost assessment | Operational Strategy: Level 4 hub-to-hub operation<br>Costs: Operational cost including hard-/software, driver, fuel, and insurance |
| Roland Berger 2016 | Operational cost assessment | Operational Strategy: Level 4 and 5 long- and short-haul operational strategies<br>Costs: Operational cost including driver, fuel, equipment, insurance, and hard-/software |
| Sharpe 2017 | Review on barriers to the market uptake of automated trucks | Costs: Breakdown of operational costs for trucks |
| Slowik 2018 | Literature review on benefits and drawbacks of automated truck adoption | Costs: Costs of components and entire automation system<br>Fuel Savings: Fuel saving strategies like platooning, cruise control, and eco-driving |
| Tsugawa 2016 | Review of truck platooning projects | Fuel Savings: Platooning studies with different constellations (vehicle number, weight, distance, speed) |

In the next chapter, we will build upon our findings from this chapter and outline potential gaps in passenger as well as freight transport research.

# 4  Requirements and challenges of modeling and analyzing transport

We will now look at the requirements of transport modeling in view of automated driving. Based on our analysis in the previous chapters, we point out what current models lack and where research gaps may exist. We will first deal with passenger transport and then go on to freight transport, and ultimately address the technical requirements and role of infrastructure.

## 4.1 Passenger transport

Passenger transport is highly complex in terms of its use cases. Modeling must consider different trip purposes (e.g., commuting, leisure, shopping, and business), different living and traffic situations (e.g., urban areas and rural areas) and, lastly, different trip lengths (e.g., traveling between cities or within them). All of these factors affect the potential use scenarios and usefulness of automated driving. To understand the specific requirements of each, we will address these cases in the following.

### 4.1.1 Commuting and business

We will start with commuting, as the objective behind these trips is quite clear: People need to get to work or back home thereafter. The objective of commuters is therefore to make their journey as pleasurable as possible, or at least reduce its unpleasantness as much as possible. This can be achieved by either shortening the trip or improving the travel circumstances. The first option is only possible to a limited degree, i.e., by moving closer to the workplace or increasing traffic efficiency and speed. Increased traffic efficiency is thought to be achievable through a high share of automated vehicles in the transport network. Otherwise, the focus for automated vehicles is on improving the quality of the journey.

In 2019, about 59.5% of all working people in Germany were commuting, i.e., working in another municipality than the one in which they were living (BMI, 2022). The average commuting distance is 16.9 km (BMI, 2022). Most people use a car for their commute, totaling 68% in 2020 (Destasis, 2022).

We expect that automation levels 1–3 will not offer significant benefits to the driver. Therefore, we assume that levels 1–3 of automation will only reduce stress. Levels 4 and 5, however, bring more benefits to commuting by car. At these two levels, the driver is freed of all driving tasks and does not need to be attentive to it at all. Besides making driving even more relaxed, this opens completely new usage scenarios for in-vehicle time. People could pursue leisure activities like reading, watching television, or even sleeping. Another option would be to work during the commute, which could either be used to work more or lead to a reduction in the time that must be spent at the workplace, which would therefore offer more time at home for personal activities.

Business trips are similar to commuting trips. The purpose of the trip is to perform work at the destination. The main differences between them are in the total daily mileage. Typical daily travel distances during business trips might be up to three times the distance of typical commuting days (Nobis and Kuhnimhof, 2018). Relieving the driver from the driving task would therefore be more favorable. Furthermore, the option to perform work-related tasks during the trip might be higher during these.

### 4.1.2 Leisure trips

For leisure trips, there are two different scenarios to consider. In the first case, people want to get to a destination where they conduct their leisure activity, e.g., engaging in sports or having dinner with friends. In this case, traveling to the destination is only a means to the end and can be considered equal to the case of commuting. In addition to that, leisure activities become more accessible (Rojas-Rueda et al., 2020), as people (especially young children or elderly people) are no longer dependent on having a driver who can take them to the locations of their desired activities. Participation in such activities might therefore increase and further transport demand be induced. Nevertheless, as already mentioned, knowledge about potential new user groups remains limited (Krueger et al., 2016). The second case to consider is when the trip is already part of the leisure activity. For example, one could be driving to an event with one's friends, time spent on this journey will be perceived as more pleasurable, and the improvements that can be achieved through automated driving will therefore be more modest.

Holiday trips can be regarded as long-range versions of leisure trips. Nevertheless, the longer travel distance brings further benefits to them. During holiday trips, oftentimes a break must be taken for the driver to rest, in addition to refueling and maybe food breaks. With automation, this break is no longer needed so the trip can be performed faster. Additionally, there is the option of shifting holiday trips to night-time periods, during which the passengers can sleep. This could massively reduce the importance of travel time for the trip. Long-distance car trips could be favored against flights with the benefits of taking more luggage and traveling more comfortably.

### 4.1.3 Shopping trips

Shopping trips have a typical distance of around 5 km in Germany and are therefore much shorter than commuting and leisure trips, with a typical distance of 15–16 km (Nobis and Kuhnimhof, 2018). Therefore, the benefits of automated driving to make the time spent in the vehicle more pleasurable might be negligible. There is, however, a large benefit to shopping trips. With automated vehicles, there is no need to search for a parking space, which can (depending on where one lives) take up several minutes (Hagen et al., 2021), walking to and from the car not included.

For the modeling of passenger transport, this means that automation must be represented via changes in various aspects. First, the value of travel time must reflect new time usage options and might therefore decrease, which could lead to a shift in modal preferences in favor of automation. Second, the trip duration of automated vehicle trips must be adopted. On the one hand, door-to-door trips that obviate the need to search for a parking space save time for searching and walking. On the other hand, taking longer/slower routes to avoid high traffic volumes prolongs individual trips in favor of overall traffic efficiency. Third, trips should be modeled with flexible departure times. Commuting trips could be shifted by minutes up to an hour and holiday trips might be shifted to nighttime.

When assessing and modeling the potential impacts of automated driving, many of the studies presented in chapter 3.1 assume the benefits of automated driving, expressed by a decreased value of travel time, to be high, i.e., a 50% reduction assumed by Auld et al. (2018) and Correia and van Arem (2016), as well as a 25–75% reduction assumed by Auld et al. (2017). This is not in line with the presented results of time usage surveys, where it has been found that people do not yet perceive the benefits of automated driving (Yap et al., 2016). Researchers may believe that people want to use their time as productively as possible and thereby overestimate the benefits of automated vehicles (Rashidi et al., 2020). In contrast, assumptions should reflect the actual preferences and the impact of different magnitudes in the assumptions should be analyzed, as model assumptions were found to be highly relevant for their results (Soteropoulos et al., 2019).

In addition, the prospects of shared automated vehicles may have been affected by the Covid-19 pandemic and their market uptake might have been hindered by it. During the pandemic, the preference for private vehicle trips over public transport increased (Kolarova et al., 2021). Furthermore, the use of shared vehicles declined, as people traveled less, had health concerns or wanted to avoid shared environments (Loa et al., 2022). Most people stated that they would substitute shared vehicle trips with private vehicle trips, either driving themselves or being driven by someone they knew (Loa et al., 2022). This rise in private vehicle trips could lead to increased traffic and higher overall vehicle kilometers travelled, and is to be seen critically from a sustainability perspective (Kolarova et al., 2021), as people might stick to these behaviors even after the pandemic (Sunder et al., 2021). Mode choice studies must consider this behavioral shift an obstacle to shared vehicle adoption and adjust the preferences used. Furthermore, in future modeling studies, the risks of relying on mobility modes that increase direct (sharing rides) and indirect (sharing vehicles) contact between people should be analyzed against the threat of another pandemic.

In contrast to the current approaches described in chapter 3.1, when modeling the impacts of automated driving, the entire transport system must be considered. Especially when new transport concepts like car- and ride-sharing come into play, the ways these might compete with other modes like public transport and in which ways they might complement them must be investigated. Modeling automated vehicles separately from other transport modes is not sufficient. Furthermore, instead of focusing on urban areas and commuting trips, all region types of a country and all trip purposes must be considered in the analyses. As a last point, market introduction scenarios with deployment plans are needed when assessing the impact of automated driving with respect to the sustainability goals in the transport sector.

## 4.2   Freight transport

As described in chapter 3.2, freight transport research does not need to deal with as many possible scenarios as passenger transport. Current research is dealing with cost-saving opportunities for automated trucks in the areas of driver costs and platooning. Besides full automation, partial automation (motorway driving) and corresponding operational strategies are considered. Research therefore already covers most aspects. Nevertheless, we wish to point out some remaining issues and gaps.

Current studies on the economic feasibility of automated freight transport mostly focus on selected use cases for freight operation in the United States, but economic feasibility in other countries might be quite different after all. The transport distances of road freight transport in countries like Germany might be much shorter than the distances stated in the studies. The average road freight transport distance for consumer goods in Germany in the year 2020 was around 150 km (KBA, 2021), which is just 18.8% the length of the freight trip example noted in chapter 3.2 for US trucking operations (Roland Berger, 2018). The share of highway driving and consequently the expected initial benefits of mixed automated and human driving might therefore be much smaller for German freight operations. It is therefore an important task for research to evaluate the potential for automated trucks for a wide variety of use cases with different driving distances and speeds.

Changes in truck operation, e.g., the potential for fuel consumption reductions, must be analyzed in greater detail. As mentioned in chapter 3.2, platooning has already been investigated in this context quite extensively and driving speed reductions became a research interest lately.

Another benefit of automated trucks relating to a higher temporal flexibility is the potential to reduce the number of freight transport trips during peak traffic hours. In these time periods, significantly increased passenger car traffic often leads to congestion. By reducing truck-based freight traffic, more road capacity would be reserved for cars, resulting in reduced congestion and more efficient traffic. Future analyses should therefore focus on road network utilization and freight transport operations to determine the potential for shifting freight transport over time.

A further point to consider for automated freight transport is the use of different drivetrains. With respect to sustainable transport, battery- and fuel cell-electric concepts are increasingly attractive. Automated freight transport, i.e., not being dependent on the driving hours of a human driver, provides more flexibility for vehicle recharging and refueling at cost-optimal times and locations. Freight transport should therefore be modeled in relation to the energy system in order to analyze potential synergies.

Ultimately, as already noted in chapter 3.2, strong cost reductions in road freight transport might lead to a shift in goods from rail freight to road freight transport. Furthermore, shorter delivery times might also lead to a shift toward road freight transport and increases in road transport capacity. These effects must be considered; automated road freight transport cannot be considered on its own. Modeling studies must include a mode choice option for freight operations that compares costs for the different transportation modes in order to fully assess the effect of automated road transport.

## 4.3  Technical requirements

We now consider the technical requirements of automated driving laid out in chapter 2 and how to implement these in modeling approaches. We start with requirements on individual vehicles and afterwards turn to infrastructure requirements.

### 4.3.1  Individual vehicles

Common practice for automated vehicles is that they rely on their own sensors and can function for themselves without depending on other vehicles or infrastructure. This is required in order for the vehicles to be able to drive anywhere. For a first deployment of automated vehicles, it is therefore sufficient to look at the vehicle technology on its own. In modeling studies, the technology will be implemented as additional costs for the automated vehicles. Drawing on our remarks in chapter 2, we can compile cost estimates for the automated vehicle technology. Taking eight cameras for 4,000 USD, two radar sensors for 6,000 USD, as well as one large and two small lidar sensors for 10,000 USD, we end up at costs of 20,000 USD for the vehicle sensors. In addition to that, there are costs for the computational hardware and the automated driving software. The software might be the most critical part to assess. Its price (in the form of automated vehicles) is dependent on the expected benefits from users. Cost estimates for automated trucks from Roland Berger predict software costs of 20,000 USD (Roland Berger, 2016) and 5,000 USD (Roland Berger, 2018). We assume a value of 15,000 USD and add another 5,000 USD for the computational hardware, so we gain another 20,000 USD for the computational system. In total, we arrive at additional costs of 40,000 USD for automated vehicles. In future modeling studies, the different costs for automated vehicles should be assumed and the impact of these on the total cost of ownership and vehicle market uptake analyzed.

To see what impact this can have on freight operations, Figure 7 shows a simple cost comparison of conventional and automated semi-trucks, based on the following assumptions: The only differences between these two modes are a 40,000 EUR higher purchasing cost and the omission of driver costs for automated semi-trucks; the semi-trucks have a yearly mileage of 114,000 km (Wietschel et al., 2017); driver wages are €0.3333/km (€20/h driver wages based on €2825/month (Federal Office for Statistics, 2022) with 170 h/month and 20% extra gross for employers with an average driving speed of 60 km/h); no interest rate is considered. Besides the conventional semi-trucks, three scenarios are depicted, with 20%, 50%, and 100% of the distance driven being automated. For the case of 20% automation, the upfront extra costs for the automation system are nearly compensated for by savings in driver wages after a five-year period, and even earlier for the other two scenarios.

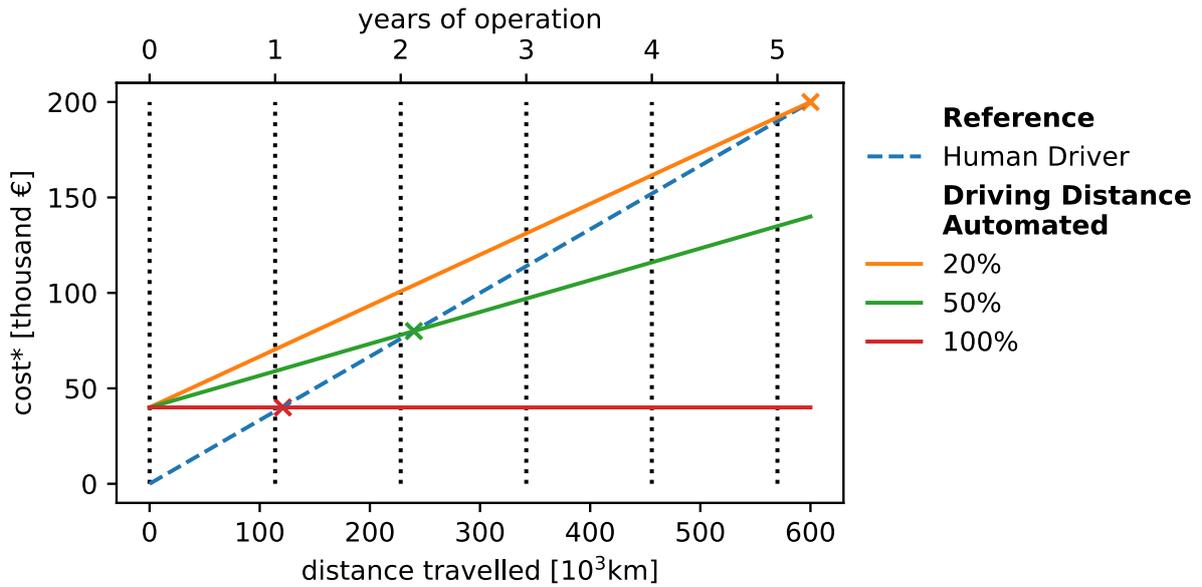

*Figure 7. Cost comparison for a human-driven and automated semi-truck. The crosses mark the breakeven of the automated driving strategies after 120, 240, and 600 thousand km respectively. 114,000 km are assumed as yearly mileage. \*Only additional purchasing and driver costs are considered, as all other costs are assumed to be equal for the two modes.*

For a second comparison, we use the total cost of ownership model from Kraus et al. (2021). We explore the costs for a fuel cell-electric semi-truck built in the year 2030 with a yearly mileage of 100,000 km. The difference in purchasing costs between the automated and non-automated vehicle are 20,000 EUR. Driver costs are €0.3333/km, as before. Finally, we consider a lifetime of 11 years for conventional semi-trucks and five years for automated ones due to rapid technological developments in automation systems. Figure 8 presents the total cost of ownership per vehicle kilometer for a human-driven vehicle and an automated truck against different shares of automated driving. We find the TCO per kilometer to be equal when 50% of the driving distance is automated.

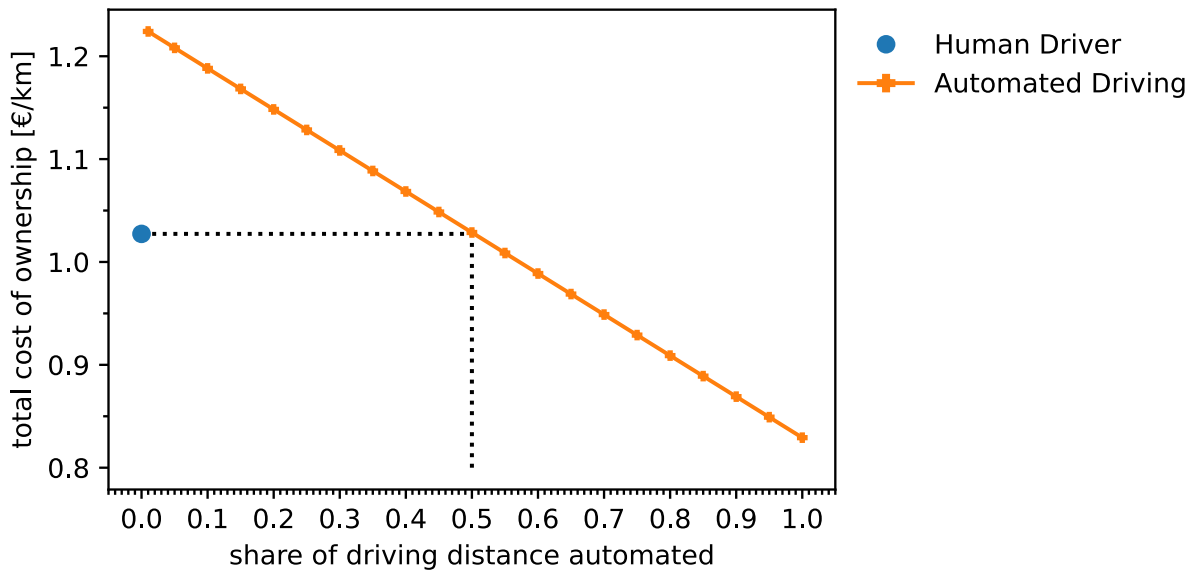

*Figure 8. Total cost of ownership comparison for a human-driven and automated fuel cell-electric semi-truck. Own calculation based on (Kraus et al., 2021).*

### 4.3.2   Infrastructure

After considering the individual vehicle perspective, Figure 9 summarizes the role of infrastructure for automated driving. In the following, we will first consider requirements for the infrastructure before moving on to potential benefits. Most importantly, the vehicle must be able to perceive infrastructure (road markings, traffic signs and traffic lights) correctly. Small changes in traffic signs, e.g., putting a sticker on them, can make them unreadable to automated vehicles, even though this might seem like a minor issue for human drivers (Hancock et al., 2019; Martínez-Díaz and Soriguera, 2018). Furthermore, infrastructure must be consistent across regions and borders (Martínez-Díaz and Soriguera, 2018). Building smart infrastructure and enabling vehicle-to-infrastructure (V2I) communication could foster some further positive effects. On an individual level, adding landmarks or electronic signatures to the infrastructure could enhance automated vehicles' location algorithms. On a systemic level, if the automated vehicles know, for example, how fast they should go to be able to pass the next intersection, they can adapt their driving speed and adopt a more efficient driving style. Furthermore, automated vehicle routing can be optimized based on information about traffic volumes in specific areas. As described in chapter 2.2, the cost-optimal placement of infrastructure communication technology (roadside units) has already been modeled under latency and coverage restrictions. The costs of placement must be further compared against the potential benefits of the technologies. In this context, analyses of traffic efficiency gains might easily be based on costs, whereas safety gains could be harder to quantify when injuries and the death of people are the object under investigation.

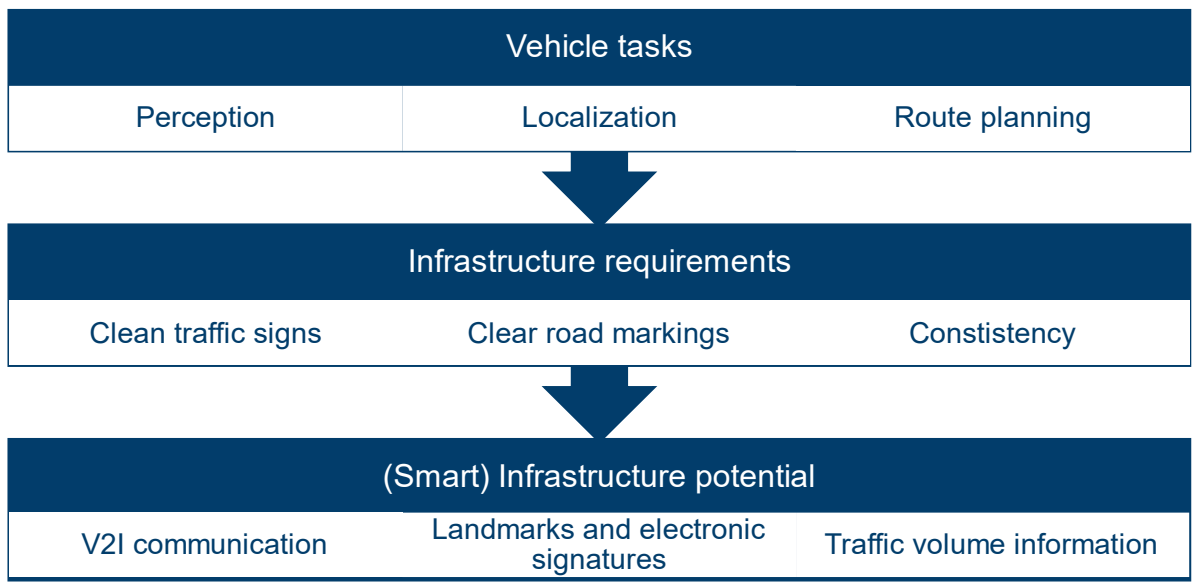

*Figure 9. Infrastructure requirements and potential benefits for automated driving (based on Campbell et al. (2010), Hancock et al. (2019), Martínez-Díaz and Soriguera (2018), and Pendleton et al. (2017)).*

## 5   Proposal of an analytical framework

To answer the question: What is the impact of automated driving on the techno-economic performance transport systems? We now propose guidelines for future modeling studies in passenger and freight transport and present an analytical framework, based on the analyses performed in the previous chapters. These will cover transport modes, trip purposes, vehicle types, and considered regions from a vehicle perspective, as well as the road network utilization from a systemic one. The analytical framework is summarized and displayed in Figure 10.

Generating a reliable transport demand is the first step in the analytical framework. Transport demand needs to be created for passenger and freight transport.

For passenger transport, it is essential to assess all trip purposes in order to gain a holistic picture of transport demand. In doing so, the influence of automation must be modeled differently for the various trip purposes. This might result in different effects on trip utility. A comparison with other transport modes must be made, as changes in trip utility for automated vehicles might lead to modal shifts. As a last point, different traffic and living situations should be considered. The impact of automated vehicles might differ drastically between cities and rural areas because transport demand and available transport modes are entirely different. Therefore, regionalized agent-based models are called for to investigate future mobility behavior. An example for such a model is the transport demand model developed by Reul et al. (2021). It is an activity-based mode-choice model which simulates the behavior of 1,000 representative agents. The model is based, among other factors, on population size and distribution as well as activity availabilities and locations. The model was extended to cover the whole of Germany, spatially resolved on the level of *Gemeindeverbände* (municipal associations). It lies in between LAU1 and LAU2 regional classification and consists of 4620 regions for Germany. The model can be calibrated to real-world traffic demand and mobility behavior by adjusting the preferences of the agents. Automated vehicles, new mobility concepts, changes in the mobility behavior, and different countries could therefore be depicted well by the model.

For freight transport, operational strategies for automated trucks must be developed and their feasibility analyzed for different driving profiles of truck classes. Long-haul trucks driving on highways most of the time will presumably profit much more from the automation of highway driving than construction vehicles. Analyses also need to consider the individual benefits automation might bring to truck classes with respect to operational times. In the end, a comparison with rail and inland waterway transport must be made. Cost reductions through automation might lead to modal shifts towards road freight transport, increasing the road transport volume. The impact of automation for freight operations must be quantified by TCO models for specific use cases. In the end, a mode choice analysis of different transport modes based on the TCO must be performed and the transport demand per mode determined. An example for a TCO model is the vehicle cost model developed by Kraus et al. (2021). The model calculates the total cost of ownership for a vehicle based on manufacturing, operation, and maintenance costs. The manufacturing costs are calculated bottom up based on the components of the vehicles. Therefore, it is easy to add additional costs for the sensors and computational system of automated vehicles. Operational costs consist of fuel costs and driver costs. The operational strategies can be developed based on trip statistics of trucks collected by mobility surveys (e.g., *Kraftfahrzeugverkehr in Deutschland (KiD)* (motor vehicle traffic in Germany) (WVI, IVT, DLR, KBA, 2012) for Germany). The transport demand can be based on freight flow statistics, e.g., yearly European freight flows on NUTS3 regional classification as developed by Speth et al. (2022). The spatial and temporal resolution need to be as high as possible, which can for example be achieved by breaking down the freight flows for each NUTS3 region based on the size of the industrial sites within them and using trip starting times from transport statistics or traffic flow data from counting loops.

In order to analyze the impact and potential of automated vehicles on the traffic network, the combined transport demand of passenger and freight transport must be considered. Road capacities must be adopted to account for possible changes in traffic efficiency through automated vehicles. To assess these potentials, traffic flow simulations for mixed traffic of automated and non-automated vehicles must be performed. On this basis, road network analyses should be conducted to identify potential bottlenecks in the traffic network. Lastly, strategies for relieving the traffic network like shifting (freight) transport times can be investigated.

After generating the transport demand for passenger and freight transport it needs to be transferred to the actual road network.

As a first step, road network data is needed, which can for example be taken from OpenStreetMap (OSM). Then a link between the regional classification of the transport demand and the road network needs to be created, e.g., assigning a node in the road network to each region representing its respective midpoint. From that the actual trips in the road network can be generated. The current traffic network situation can be analyzed and potential bottlenecks identified, especially in the context of increasing future transport demands. For the assessment of changes in traffic flow caused by automated vehicles, microscopic traffic flow simulations can be performed, e.g., with the traffic simulation software SUMO, developed by the DLR (Lopez et al., 2018), by adjusting driving characteristics like following distances and reaction times to depict the behavior of automated vehicles. The insights on traffic capacity changes gained from these simulations can then be used to reevaluate the traffic network situation regarding bottlenecks. In a further step the potential influence of changing freight departure times to lower the peak traffic in the network can be investigated.

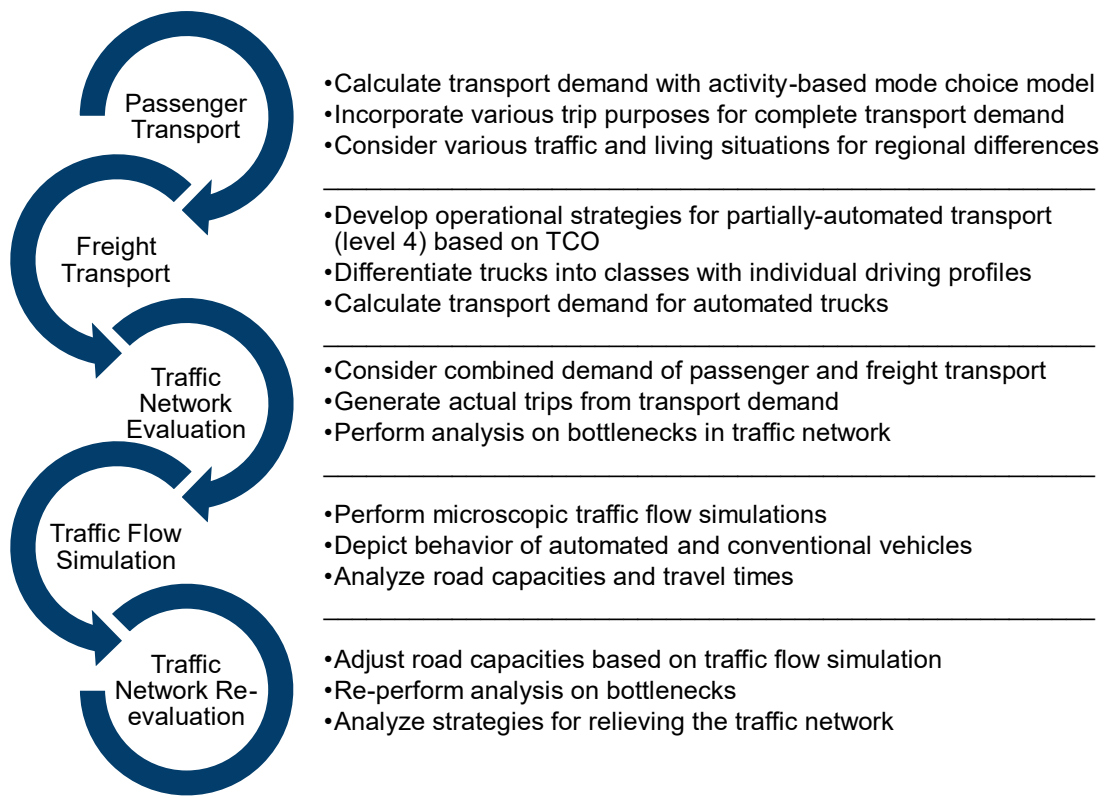

*Figure 10. Analytical framework for analyzing the impact of automated vehicles in a transport system.*

# 6  Conclusions

In this study we analyzed the state-of-the-art of automated driving technologies and reviewed current literature on the impact of automated vehicle usage in passenger and freight transport. We then pointed out areas of concern for future research.

With respect to the state-of-the-art of automated driving technology, we found that research is mainly concerned with improving environmental perception and in particular the high variety of traffic situations and environmental conditions an automated vehicle may encounter poses a major challenge. The algorithms must provide reliable results while keeping computational times and energy demand low. A further research frontier is the use of infrastructure to improve automated driving with respect to safety and traffic performance. The potential benefits of these technologies must be specified and (cost-)optimal distributions developed. We conclude that it can be expected that automated vehicles for passenger transport will not be available for the general public in most areas before 2030–2040. In the area of freight transport, on the other hand, level 4 automation for highway operations could already be commercially-available within the next few years.

For passenger transport, we found that most studies introducing shared automated vehicles focus on the impacts of automated vehicles separated from other modes of transport, in that they analyze specific (urban) regions or focus on commuting as the main trip purpose. We instead propose that the impact of automated vehicles must be studied holistically, focusing on three main points. First, mode choice must be modeled to assess the interaction of automated vehicles with other transport modes. Second, different regions, e.g., urban and rural as well as customized mobility concepts for these, must be considered. Third, the complete range of trip purposes must be modeled to properly reflect the transport demand.

With respect to freight transport, we found that most studies focus on the primary effects of automation in freight transport for fuel- and cost-saving. We propose further analyzing the usage potentials of automation in other regions with a focus on operational strategies adjusted to specific use cases. In addition to that, potential changes in operational strategies should be analyzed, including the potential of slower driving for fuel reduction and measures for improving traffic efficiency like excluding freight transport from commuting hours to relieve traffic congestion. In the end, road freight transport must be analyzed in combination with rail and other transport modes to examine potential modal shifts resulting from cost reductions or gains in delivery times and capacity increases through the automation of road freight operations.

For an analysis of the overall performance of the traffic network, passenger and freight transport demands must be combined on a high spatial and temporal resolution. The current network utilization needs to be analyzed and bottlenecks identified. For the assessment of changes in traffic flow and road capacities caused by automated vehicles, microscopic traffic flow simulations should be performed. Afterwards, the traffic network situation regarding bottlenecks can be reevaluated and the impact of automated vehicles on the traffic situation estimated.

# 7  Acknowledgments

This work was supported by the Helmholtz Association under the program, "Energy System Design."

## Apendix A

To assess the claim by Fagnant and Kockelman (2014) that vehicle production rates might remain unchanged when the vehicle fleet size decreases through a stronger utilization of the individual vehicles, we perform a brief analysis of vehicle production rate dependencies in the following.

At first, we assume that the vehicle production rate $PR_V$ is given by the vehicle fleet size $N_V$, divided by the vehicle lifetime $LT_V$ in years:

$$PR_V = \frac{N_V}{LT_V}.$$

Next, we assume that the vehicle lifetime in years $LT_V$ is given by the total mileage of a vehicle over its lifetime $LTK$, divided by its yearly mileage $VKT_{year}$:

$$PR_V = \frac{N_V \, VKT_{year}}{LTK}.$$

The product of the vehicle fleet size $N_V$ and the yearly mileage of a vehicle $VKT_{year}$ equates to the total vehicle mileage of all vehicles $VKT$:

$$PR_V = \frac{VKT}{LTK}.$$

Finally, we assume that the total vehicle mileage of all vehicles $VKT$ is given by the total passenger mileage $PKT$, divided by the average vehicle occupancy $OCC$:

$$PR_V = \frac{PKT}{LTK * OCC}.$$

The vehicle production rate needed might therefore decrease if vehicle lifetime mileages increase through electric drivetrains and with less attrition due to aging because of higher yearly mileages. This decrease might be countered if overall passenger kilometers traveled increase because traveling by car might become cheaper or otherwise more convenient. Furthermore, empty vehicle travel for the relocation of automated vehicles might also lead to lower occupancy rates, further increasing the vehicle production rate. However, this effect might be countered by increases in vehicle occupancy rates via the emergence of ride-sharing concepts. A range of scenarios for the vehicle production rate are shown in Figure 11.

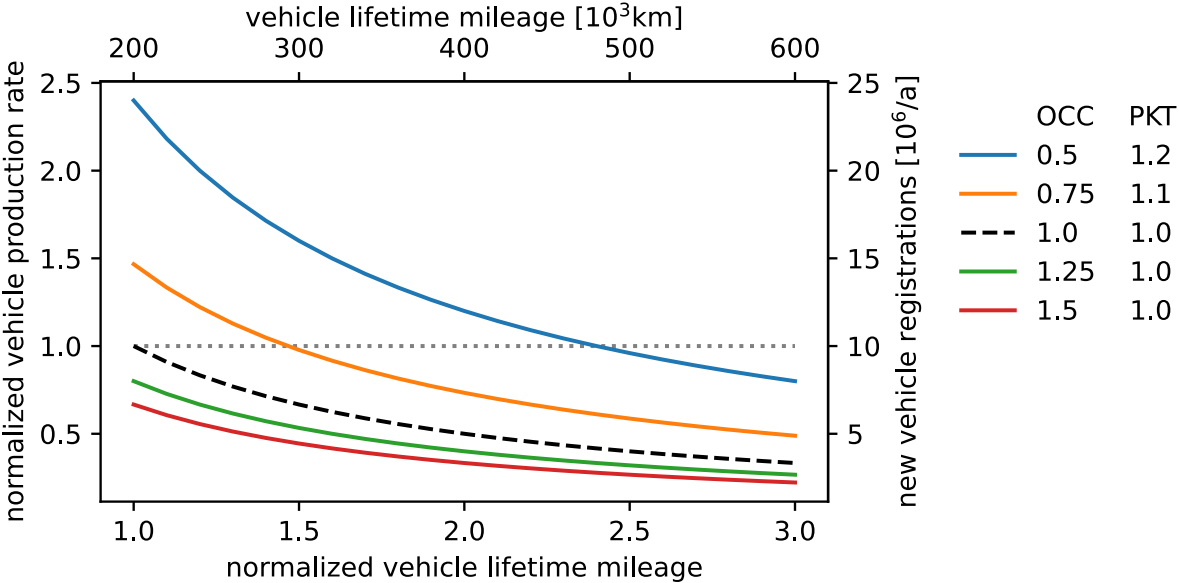

*Figure 11. Changes in the vehicle production rate in comparison to the baseline depending on changes in vehicle lifetime kilometers traveled for four scenarios of changes in occupancy rate (OCC) and passenger kilometers travelled (PKT). The black dashed line represents the case with no changes in vehicle occupancy rate and passenger kilometers traveled. New vehicle registrations per year represent values for the EU, with 10 million vehicles as a baseline for 2020 (ACEA, 2022). The baseline for vehicle lifetime mileage of 200,000 km represents the suggestion from Weymar and Finkbeiner (2016).*